\documentclass[11pt]{article}
\usepackage{amsmath,amssymb,color,graphics,epsfig}

\usepackage{amsfonts}

\newcommand{\be}{\begin{equation}}
\newcommand{\ee}{\end{equation}}
\newcommand{\bea}{\setlength\arraycolsep{2pt} \begin{eqnarray}}
\newcommand{\eea}{\end{eqnarray}}
\newcommand{\nn}{\nonumber}

\def\ft#1#2{{\textstyle{\frac{\scriptstyle #1}{\scriptstyle #2} } }}
\def\fft#1#2{{\frac{#1}{#2}}}

\def\0{{\sst{(0)}}}
\def\1{{\sst{(1)}}}
\def\2{{\sst{(2)}}}
\def\3{{\sst{(3)}}}
\def\4{{\sst{(4)}}}
\def\5{{\sst{(5)}}}
\def\6{{\sst{(6)}}}
\def\7{{\sst{(7)}}}
\def\8{{\sst{(8)}}}
\def\sst#1{{\scriptscriptstyle #1}}

\begin{document}

\begin{flushright}
\end{flushright}

\vspace{25pt}
\begin{center}
{\large {\bf Exact formation of hairy planar black holes}}

\vspace{10pt}
Zhong-Ying Fan$^{1}$ and Bin Chen$^{1,2,3}$

\vspace{10pt}
{\it $^{1}$Center for High Energy Physics, Peking University, No.5 Yiheyuan Rd,\\}
{\it  Beijing 100871, P. R. China\\}
\smallskip
{\it $^{2}$Department of Physics and State Key Laboratory of Nuclear Physics and Technology,\\}
{\it Peking University, No.5 Yiheyuan Rd, Beijing 100871, P.R. China\\}
\smallskip
{\it $^{3}$Collaborative Innovation Center of Quantum Matter, No.5 Yiheyuan Rd,\\}
{\it  Beijing 100871, P. R. China}

\vspace{40pt}

\underline{ABSTRACT}
\end{center}
We consider Einstein gravity minimally coupled to a scalar field with a given potential in general dimensions. We obtain large classes of static hairy planar black holes which are asymptotic to AdS space-times. In particular, for a special case $\mu=(n-2)/2$, we obtain new classes of exact dynamical solutions describing black holes formation. We find there are two classes of collapse solutions. The first class solutions describe the evolution start from AdS space-time with a naked singularity at the origin. The space-time is linearly unstable and evolves into stationary black hole states even under small perturbation. The second class solutions describe the space-time spontaneously evolves from AdS vacua into stationary black hole states undergoing non-linear instability. We also discuss the global properties of all these dynamical solutions.

\vfill {\footnotesize Emails: fanzhy@pku.edu.cn,\quad bchen01@pku.edu.cn}

\thispagestyle{empty}

\pagebreak

\tableofcontents
\addtocontents{toc}{\protect\setcounter{tocdepth}{2}}




\section{Introduction}
AdS/CFT correspondence provides a new approach to study non-equilibrium thermalization of strongly coupled field theories\cite{Balasubramanian:2010ce,Balasubramanian:2011ur}.
The dual picture in the gravity side is the space-time evolves from pure AdS vacua into some stationary black holes states at late times. Hence
the dynamical formation of planar black holes that are asymptotic to AdS space-times is of particular interests and importance in the field of AdS/CFT correspondence. Besides, the (in)stability of
AdS vacua is also an important subject in General Relativity. This is usually studied by both numerical and half-analytical methods in literature \cite{Bizon:2011gg,Buchel:2012uh,Wu:2013qi,Buchel:2013uba,Bhattacharyya:2009uu}. Generally speaking, AdS vacua with a massive scalar satisfying the  Breitenlohner-Freedman bound is linearly stable against small perturbation. However, numerical results also indicate that space-time can undergo non-linear instability and evolve into some final black holes states. In short, the construction of dynamical solutions in General Relativity has both applications in quantum field theories and its own interests and importance.

Unfortunately, there are few examples of exact dynamic solutions having been found for given Lagrangians. Recently, a first example of exact dynamic solutions describing black holes formation was constructed in some
minimally coupled Einstein-scalar gravity in four dimensions \cite{Zhang:2014sta,Zhang:2014dfa}. This solution was soon generalized to include electric charges \cite{Lu:2014eta}. Analogous dynamical solutions were also found in three dimensions \cite{Xu:2014xqa,Ayon-Beato:2015ada} and higher dimensions \cite{Fan:2015tua} in some non-minimally coupled Einstein-scalar gravities.

The motivation of current paper is to construct more exact dynamic solutions of black holes formation. We focus on minimally coupled Einstein-scalar gravities. We would like to construct dynamical black holes with planar topology which may have further applications in the AdS/CFT correspondence. Our strategy follows \cite{Zhang:2014sta}. We first construct static black holes with scalar hair. The solutions contain only one independent integration constant associated with the scalar field. This is a non-conserved quantity. We further promote it to be dependent on the advanced times and solve the equations of motions in Eddington-Finkelstein like coordinates. If all the equations of motions can be analytically solved with the time dependent scalar parameter satisfying a single dynamic evolution equation, we succeed in constructing exact dynamic solutions. Yet this procedure does not always work well. In fact, we cannot obtain further solutions in most cases. Fortunately, for a special parameter $\mu=(n-2)/2$, we successfully construct some new dynamical solutions in general dimensions. Their properties and global structures are also discussed with details.

This paper is organized as follows. In section 2, we construct new static hairy planar black hole solutions in certain minimally coupled Einstein-scalar theories. We analyze their global structures and derive their thermodynamic first laws. In section 3, we follow above strategy and obtain new classes of dynamical solutions. In section 4, we give a short discussion on the global properties of general dynamical solutions. In section 5, we first studied two exact examples in which the dynamic evolution equation of the scalar parameter can be analytically solved. For general case, the scalar parameter are solved in terms of an implicit function of the advanced time. We find that these dynamical solutions can be classified into two different classes and they all describe black holes formation. We conclude this paper in section 6.

\section{Static solutions}
We consider Einstein gravity minimally coupled to a scalar field whose Lagrangian density is given by
\be \mathcal{L}=R-\ft 12 (\partial \phi)^2-V(\phi) \label{lagrangian}\ee
The covariant equations of motions are
\be G_{\mu\nu}=\ft 12\partial_\mu\phi \partial_\nu \phi-\ft 12 g_{\mu\nu}\Big(\ft 12(\partial \phi)^2+V(\phi) \Big)\,,\qquad \Box\phi =\fft{\partial V}{\partial \phi} \label{eom}\ee
This model has been extensively studied in literature \cite{Fan:2015tua,Henneaux:2002wm,Martinez:2004nb,Anabalon:2012dw,Gonzalez:2013aca,Anabalon:2013sra,Acena:2013jya,Feng:2013tza,Zloshchastiev:2004ny,Fan:2015oca}. The construction method has been well demonstrated. We will omit the construction details and just give the final results. We find that for the potential given by
\bea V&=&-\ft 12 \big(\cosh{\Phi}\big)^{\ft{\mu k_0^2}{n-2}}\Big(g^2-\alpha \big(\sinh{\Phi}\big)^{\ft{n-1}{\mu}} {}_2F_1\big(\ft{n-1}{2\mu},\ft{\mu k_0^2}{4(n-2)},\ft{n+2\mu-1}{2\mu},-\sinh^2{\Phi}  \big)   \Big)\nn\\
&\times& \Big(2(n-2)(n-1)-\mu^2k_0^2\tanh^2{\Phi} \Big)-\alpha (n-2)(n-1)\big(\cosh{\Phi}\big)^{\ft{\mu k_0^2}{2(n-2)}}\big(\sinh{\Phi}\big)^{\ft{n-1}{\mu}}\,,  \label{potential}\eea
where $n$ is the space-time dimension, $g^2\,,\alpha\,,k_0$ are numeric constants and
\be \Phi=\fft{\phi}{k_0}\,,\ee
there exist exact black hole solutions. Before presenting the solutions, let us first discuss the properties of the potential. For small $\phi$, the potential can be expanded as
\be V=-(n-2)(n-1)g^2-\ft 12 \mu k_0^2\Big( (n-\mu-1)g^2+\ft{2\mu^2}{n+2\mu-1}\alpha\, \Phi^{\fft{n-1}{\mu}}  \Big)\Phi^2+\cdots\,.\ee
We can read off the scalar mass square $m^2=-\mu(n-\mu-1) g^2$. For $0<\mu<n-1$, the mass square becomes negative and we shall require that it is above the BF bound, namely
$m^2\geq m^2_{BF}=-\ft 14(n-1)^2 g^2$. Interestingly, we find that this condition is always satisfied and $m^2=m^2_{BF}=-\mu^2 g^2$ when $\mu=\ft 12 (n-1)$. This implies that
the AdS vacua is stable against small perturbation at linear level.

Given this potential, our minimal theory (\ref{lagrangian}) admits exact planar black holes solutions with scalar hair
\bea &&ds^2=-fdt^2+\fft{\sigma^2dr^2}{f}+r^2d\vec{x}^2\,,\qquad\quad \phi=k_0\,\mathrm{arcsinh}(\ft{q^\mu}{r^\mu})\,,\nn\\
      &&\sigma=\Big(1+\ft{q^{2\mu}}{r^{2\mu}} \Big)^{-\ft{\mu k_0^2}{4(n-2)}}\,,\qquad f=g^2 r^2-\ft{\alpha\, q^{n-1}}{r^{n-3}}
      {}_2F_1\Big(\ft{n-1}{2\mu},\ft{\mu k_0^2}{4(n-2)},\ft{n+2\mu-1}{2\mu},-\ft{q^{2\mu}}{r^{2\mu}} \Big)   \,,\label{staticsol}\eea
where $d\vec{x}^2$ is the $(n-2)$ dimensional space with planar topology and $q$ is the only independent integration constant associated with the scalar. It is easy to see that the AdS vacuum emerges for vanishing $q$.

We now turn to discuss the global properties of these solutions. Asymptotically, the metric function $f$ behaves as
\be f=g^2 r^2-\ft{\alpha\, q^{n-1}}{r^{n-3}}\Big(1-\ft{\mu k_0^2(n-1)}{4(n-2)(n+2\mu-1)}\Big(\fft{q}{r}\Big)^{2\mu}
+\ft{\mu k_0^2(n-1)\big(\mu k_0^2+4(n-2)\big)}{32(n-2)^2(n+4\mu-1)}\Big(\fft{q}{r}\Big)^{4\mu}+\cdots  \Big) \,.\ee
The mass of the black holes can be read off by
\be M=\fft{(n-2)\alpha q^{n-1}}{16\pi} \label{staticmass}\,.\ee
The positiveness of the mass requires $\alpha>0$. To establish that there exists a horizon, we first note that the hypergeometric function in the metric function $f$ is positive definite due to following identities
\be {}_2F_1\Big(\ft{n-1}{2\mu},\ft{\mu k_0^2}{4(n-2)},\ft{n+2\mu-1}{2\mu},-\ft{q^{2\mu}}{r^{2\mu}} \Big)\equiv \Big(1+\ft{q^{2\mu}}{r^{2\mu}} \Big)^{-\fft{\mu k_0^2}{4(n-2)}}{}_2F_1\Big(\ft{\mu k_0^2}{4(n-2)},1,\ft{n+2\mu-1}{2\mu},\ft{q^{2\mu}}{r^{2\mu}+q^{2\mu}} \Big) \,.\ee
Therefore, for sufficient large $\alpha$, there does exist a horizon $r_0>0$ at which $f(r_0)=0$. There are vast examples which satisfy this condition and we shall not list them case by case. In fact,
a sufficient condition for the horizon is $f(r)$ is negative in the $r\rightarrow 0$ limit. We find
\footnote{$\tilde{g}^2=g^2-\alpha\,\Gamma\Big(\ft{n+2\mu-1}{2\mu} \Big)\Gamma\Big(\ft{\mu^2k_0^2-2(n-2)(n-1)}{4(n-2)\mu} \Big)/\Gamma\Big(\ft{\mu k_0^2}{4(n-2)} \Big)$.}
\be f(r)=\tilde{g}^2 r^2-\ft{2(n-2)(n-1)\alpha\,q^2}{2(n-2)(n-1)-\mu^2 k_0^2}\Big(\fft{q}{r} \Big)^{\fft{2(n-3)(n-2)-\mu^2k_0^2}{2(n-2)}}+\cdots \,.\ee
The specific form of $\tilde{g}^2$ is irrelevant in our discussion. It is clear that when $\mu^2 k_0^2<2(n-2)(n-1)$, $f(r)$ will be negative in the $r\rightarrow 0$ limit for generic $\alpha$.

The temperature and entropy density are given by
\be T=\fft{(n-1)\alpha q^{n-1}}{4\pi r_0^{n-2}}\,,\qquad S=\ft 14 r_0^{n-2}\,. \ee
The equation $f(r_0)=0$ implies that $r_0/q$ is a pure number, namely
\be dr_0=\ft{r_0}{q}dq \,.\ee
It is straightforward to derive the thermodynamical first laws
\be dM=T dS \,,\ee
and the generalized Smarr formula \cite{Liu:2015tqa,Feng:2015oea}.
\be M=\fft{n-2}{n-1}T S \,.\ee

\section{Dynamical solutions}
For a special value of $\mu$: $\mu=\ft{n-2}{2}$, we find that our minimal theory also admits exact dynamic solutions. The scalar potential turns out to be
\bea V&=&-\ft 12 \big(\cosh{\Phi}\big)^{\ft{ k_0^2}{2}}\Big(g^2-\alpha \big(\sinh{\Phi}\big)^{\ft{2(n-1)}{n-2}} {}_2F_1\big(\ft{k_0^2}{8},\ft{n-1 }{n-2},\ft{2n-3}{n-2},-\sinh^2{\Phi}  \big)   \Big)\\
&\times &\Big(2(n-2)(n-1)-\ft 14 k_0^2(n-2)^2\tanh^2{\Phi} \Big)-\alpha (n-2)(n-1)\big(\cosh{\Phi}\big)^{\ft{k_0^2}{4}}\big(\sinh{\Phi}\big)^{\ft{2(n-1)}{n-2}}\,,  \nn\label{dypotential}\eea
Its small $\phi$ expansion is given by
\bea V&=&-(n-2)(n-1)g^2-\ft 18 n(n-2)k_0^2\Big(g^2+\ft{(n-2)^2}{n(2n-3)}\alpha \Phi^{\ft{2(n-1)}{n-2}}\Big)\Phi^2-\ft{1}{96}(n-2)k_0^2 \nn\\
     &\times& \Big(3k_0^2+4(n-3)\Big)\Big(g^2+\ft{(n-2)\big(3(n-2)(5n-9)k_0^2-8(2n-3)(3n-7)  \big)}{2(2n-3)(2n-5)\big(3k_0^2+4(n-3)\big)}\alpha  \Phi^{\ft{2(n-1)}{n-2}} \Big)\Phi^4+\cdots \,.\eea
It follows that the scalar mass square is $m^2=-\ft 14 n(n-2)g^2$ which is above the BF bound.

In Eddington-Finkelstein like coordinates, our new dynamical solutions can be cast into the form of:
\bea ds^2&=&-fdu^2+2\sigma dr du+r^2 d\vec{x}^2\,,\qquad \phi=k_0 \mathrm{arcsinh}\Big(\fft{a}{r} \Big)^{\fft{n-2}{2}}\,,\\
      \sigma&=&\Big(1+\ft{a^{n-2}}{r^{n-2}} \Big)^{-\ft{k_0^2}{8}}\,,\qquad f=g^2 r^2-2\sigma \dot{a}\Big(\fft{a}{r}\Big)^{n-3}-\alpha a^2\Big(\fft{a}{r}\Big)^{n-3}
      {}_2F_1\Big(\ft{k_0^2}{8},\ft{n-1}{n-2},\ft{2n-3}{n-2},-\ft{a^{n-2}}{r^{n-2}} \Big) \nn\label{dysol}\eea
where $a\equiv a(u)$ is a single function of the advanced time $u$ and $\dot{a}$ denotes the derivative with respect to $u$. Moreover, $a(u)$ satisfies
\be 8a\ddot{a}+\Big((8-k_0^2)(n-2)-8\Big)\dot{a}^2+4\alpha(n-1)a^2\dot{a}=0 \,.\label{aeq}\ee
or more suggestively
\be \fft{\ddot{a}}{a^2}+\gamma\fft{\dot{a}^2}{a^3}+\tilde{\alpha}\fft{\dot{a}}{a}=0\,,\label{aeq2}\ee
where
\be  \gamma=(1-\ft{k_0^2}{8})(n-2)-1\,,\qquad \tilde{\alpha}\equiv \ft 12 \alpha(n-1) \,. \ee
In the asymptotical limit $r\rightarrow \infty$, the metric function $f$ behaves as:
\be f=g^2 r^2-\fft{(2\dot{a}+\alpha a^2)a^{n-3}}{r^{n-3}}+\cdots\,.\ee
The effective Vaidya mass can be read off by
\be M(u)=\fft{n-2}{16\pi}\big(2\dot{a}+\alpha a^2\big)a^{n-3} \,.\label{vaidyamass}\ee
Its time derivative is given by
\be \dot{M}(u)=\fft{(n-2)^2\,k_0^2\,\dot{a}^2a^{n-4} }{64\pi} \,,\label{massdot}\ee
where we have imposed the dynamical evolution equation Eq.(\ref{aeq2}) to cancel $\ddot{a}$ term in the final result. It is immediately seen that $\dot{M}(u)\geq 0$ since $a\geq 0$ is required in our new solutions to guarantee the reality of the scalar field for general dimensions.
The equality is taken at two points $a=0$ and $a=q$. Thus the Vaidya mass monotonically increases as the advanced time $u$ increases. Moreover, the collapse solutions should satisfy $\dot{a}\geq 0$, implying that $a(u)$ also monotonically increases during the black holes formation. Therefore, without giving the specific form of $a(u)$, the physical picture behind our new solutions is simple: the dynamical evolution of space-times is driven by the scalar; it starts from AdS space-time\footnote{The space-time does not always become pure AdS vacua at this point, though the scalar vanishes. For some parameters, the space-time can also have a singularity at the origin. For details, see section 5.} with $a(u)=0$; As $a(u)$ increases, the space-time begins collapse, resulting to the monotone increasing of the black hole mass; In the end, as $u\rightarrow \infty$, $a(u)$ approaches its equilibrium value $q$; Correspondingly, the space-time arrives at some final black hole states.

We will provide some analytical examples in which $a(u)$ is solved into closed forms in section 5. Before this, in order to study the global structures of our new solutions, let us first discuss the global properties of general dynamical solutions.

\section{Global properties of dynamical solutions}
In this section, we will give some preliminaries for studying the global properties of general dynamical solutions. The metric ansatz is given by
\be ds^2=-fdu^2+2\sigma du dr+r^2 d\vec{x}^2\,, \ee
where both $f$ and $\sigma$ are functions of $r\,,u$. Note that the metric function $\sigma$ should be positive definite outside the apparent horizon since $u$ was interpreted as the advanced time coordinate\footnote{The overall sign of $\sigma$ is a pure gauge and can be absorbed by refining the time coordinate $u$. The crucial fact is that it should be regular everywhere in the space-time except the intrinsic singularity at the origin.}. Throughout this section, we will not specify the metric functions and all the discussions are valid
for this type metric.

\subsection{Apparent horizon}
By definition, apparent horizons are governed by the sign of the expansion of radial null geodesic congruences whose tangent vector are given by:
\be k_{\pm}^a\equiv(\ft{\partial}{\partial \lambda_{\pm}})^a=\pm\Big((\ft{\partial}{\partial u})^a+\ft{f}{2\sigma}(\ft{\partial}{\partial r})^a \Big) \,,\ee
where the null geodesics are parameterized by $\lambda_\pm$ (but they may not be their affine parameters) and the ``$\pm$" signs are dual to outgoing and ingoing null geodesics, respectively.
It is straightforward to verify that $k_\pm^a$ are null and they satisfy the null geodesic equations of motions
\be k_{\pm}^b\triangledown_b k_{\pm}^a=\alpha_{\pm} k_{\pm}^a\,,\qquad \alpha_{\pm}=\pm (\dot{\sigma}+\ft 12f')/\sigma \,,\ee
where the prime and dot denote the derivative with respect to $r$ and $u$ respectively.
We find
\be \theta_\pm=g_{ab}\triangledown^bk_\pm^a-\alpha_\pm=\pm\fft{(n-2)f}{2r\sigma}\,.\label{expansion} \ee
By definition, the space-time region enclosed by the apparent horizon is specified by $\theta_+\leq 0$ where the equality defines the location of the apparent horizon. In other words, the apparent horizon is determined by
\be f\big(r_{\mathrm{AH}}(u),u \big)=0 \,.\label{apparent}\ee
From Eq.(\ref{expansion}), it is clear that outside the apparent horizon the expansion of the ingoing null geodesics $\theta_-$ is negative. This implies that the apparent horizon is future
\footnote{The existence of the event horizon is guaranteed by the existence of future outer apparent horizons. The apparent horizon is called outer, if the Lie derivative $k_-^b\triangledown_b\theta_+$ is also negative outside the apparent horizon. We have also verified that our static solutions satisfy this condition. }\cite{Hayward1,Hayward2,Ida,Wang:2003bt}.

\subsection{Local event horizon}
In general, unlike the apparent horizon, the equation for the event horizon cannot be solved in dynamical space-times, though it is globally well defined\footnote{In stationary space-time, the event horizon was equivalent to the Killing horizon which corresponds to a null Killing vector with non-vanishing surface gravity.}.
We shall adopt a local definition for the event horizon which was first introduced in \cite{Zhao:1992ad,Zhao:1995jr}. The event horizon is locally defined by a null hypersurface which preserves the intrinsic symmetry of the space-time. The resulting equation for the event horizon is very simple
\be \fft{dr_{\mathrm{EH}}}{du}=\fft{f}{2\sigma} \,.\label{event}\ee
However, we are aware of that it is not clear what the relation is between this locally defined event horizon and the ``true" event horizon. To avoid confusion, we will call it ``local event horizon". It is an effective conception for the purpose of describing the formation of black holes in dynamical space-times. Indeed, we find that this definition is reasonable for our solutions. The local event horizon always encloses the apparent horizon\footnote{This is ensured in our numerics but we do not know how to prove it in a general manner.} and it approaches the event horizon in the static limit.

\section{Examples}
Now we present some examples in which the dynamical evolution equation Eq.(\ref{aeq}) can be exactly solved. We will also analyse the global structures of these dynamical solutions.

\subsection{Case 1: $k_0= 2\sqrt{\ft{2(n-3)}{n-2}}$}
When $k_0$ taking this special value, $\gamma$ vanishes and the $\dot{a}^2$ term in Eq.(\ref{aeq2}) is cancelled. The simplified equation is
\be \ddot{a}+\tilde{\alpha}a \dot{a}=0 \,,\ee
which can be immediately solved by
\be a(u)=q \tanh{\big(\ft 12\tilde{\alpha}q u\big)}\,,  \label{sol1}\ee
We see that $a(u)=0$ at $u=0$ which is the starting point of the space-time collapse. However, this is not a stationary point since
\be \dot{a}(u)=\ft 12\tilde{\alpha}q^2\,\mathrm{sech}^2(\ft 12 \tilde{\alpha}q u)\,, \ee
 is non-vanishing at $u=0$. Hence, the space-time is unstable against perturbation at linear level. Once $u$ slightly deviates from this point, $a(u)$ will be non-vanishing and the collapse happens inevitably no matter how small $a(u)$ is. However, we should point out that the space-time is not pure AdS vacua at $u=0$, though $a(u)$ vanishes. Actually the global structure of the space-time around this point is very subtle. Without loss of generality, we focus on the  $n=4$ dimension. Around $u=0$, the scalar and the metric functions behave as
\be \phi=\fft{\tilde{\alpha}q^2 u}{r}+\mathcal{O}(u^3)\,,\quad f=g^2r^2-\fft{\tilde{\alpha}^2 q^4 u}{2r}+\mathcal{O}(u^3)\,,\quad \sigma=1-\fft{\tilde{\alpha}^2q^4u^2}{8r^2}+\mathcal{O}(u^4)\,. \ee

\begin{figure}[ht]
\begin{center}
\includegraphics[width=230pt]{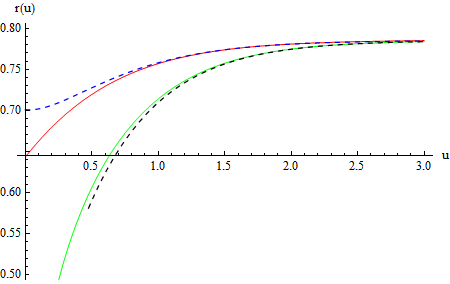}
\end{center}
\caption{{\it The evolution of local event horizon (the red line) and apparent horizon (the green line) for $n=4$ dimension. Near the static limit, both the horizons approach the event horizon exponentially fast. As $u\rightarrow \infty$, they can be solved in terms of Taylor series of $e^{-\tilde{\alpha}q u}$, which are plotted as dashed lines (to third order). Some constants $g^2\,,\tilde{\alpha}\,,q$ have been set to unity.}}
\label{fig1}\end{figure}
Some curvature invariants are given by
\bea &&R^{\mu}_{\mu}=-12g^2-\fft{\tilde{\alpha}^2q^4 u}{r^3}+\mathcal{O}(u^2)\,,\quad R_{\mu\nu}R^{\mu\nu}=36g^4+\fft{6g^2\tilde{\alpha}^2 q^4 u}{r^3}+\mathcal{O}(u^2)\nn\\ &&R_{\mu\nu\lambda\rho}R^{\mu\nu\lambda\rho}=24g^4+\fft{4g^2\tilde{\alpha}^2q^4u}{r^3}+\mathcal{O}(u^2)\,. \eea
It is easy to see that the space-time becomes pure AdS at $u=0$ except the $r=0$ point. At $u=0\,,r=0$ point, the curvature polynomials are not well defined and the existence of the singularity strongly depends on the path to this point on the $(u\,,r)$ plane. For some paths such as $r\propto u^\delta$ with $\delta\leq 1/3$, the singularity disappears. However, we find that all these trajectories are space-like, indicating that this singularity is non-traversable. Nonetheless, we can still connect the space-time to an AdS vacuum for $u<0$. The birth of the evolving black hole can be understood as a vacuum being kick-started by a naked singularity\footnote{Certainly, the energy-momentum tensor is also singular at this point. By ``naked", we mean the apparent horizon vanishes in $u\rightarrow 0$ limit. This is ensured in our numerics. We can also verify this half-analytically by developing Taylor series around the $u=0$ point.}.

In Fig.\ref{fig1}, we plot the local event horizon and apparent horizon as a function of the advanced time $u$ during the formation. The apparent horizon is always inside the local event horizon and both of them approach the event horizon in the static limit. As $u\rightarrow \infty$, the system reaches the static limit exponentially fast. We find
\be a(u)=q\Big(1-2e^{-\tilde{\alpha}q u}+\cdots\Big)\,,\qquad M(u)=M_0\Big(1-2(n-1)(n-3)e^{-2\tilde{\alpha}qu}+\cdots \Big) \,,\ee
where $M_0$ is the static black hole mass given by (\ref{staticmass}). Hence, we can define a characteristic relaxation time
$u_0=1/(\tilde{\alpha}q)\sim 1/M_0^{\fft{1}{n-1}}$. It is clear that for a bigger mass $M_0$ the relaxation time becomes shorter.
This is characteristic for our solutions. At large $u$, the evolution equations for the local event horizon and apparent horizon can also be solved in terms of $e^{-\tilde{\alpha} q u}$ series. We find
\be r_{\mathrm{EH}}=r_0-r_1 e^{-\tilde{\alpha}q u}-r_2 e^{-2\tilde{\alpha}q u}+\cdots\,,\qquad
r_{\mathrm{AH}}=r_0-\tilde{r}_1 e^{-\tilde{\alpha}q u}-\tilde{r}_2 e^{-2\tilde{\alpha}q u}+\cdots\,,\label{horizon}\ee
where $r_0$ is the event horizon of the static black hole; the series coefficients $r_i\,,\tilde{r}_i$ are solved by $n\,,q\,,r_0$, resulting to lengthy expressions which are not suggestive to give. In Fig.\ref{fig1}, we plot these series as dashed lines. We see that they perfectly coincide with the numerical solutions at large $u$ region.

\begin{figure}[ht]
\begin{center}
\includegraphics[width=230pt]{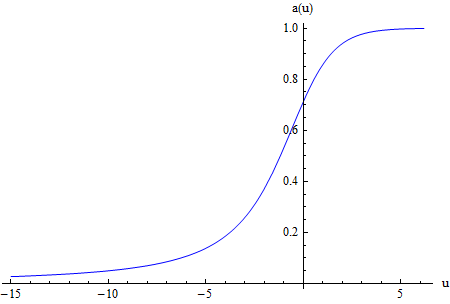}
\end{center}
\caption{{\it The plot of $a(u)$ as a function of $u$. As $u$ runs over $(-\infty\,,\infty)$, $a(u)$ evolves from one stationary point $a=0$ (corresponding to the AdS vacuum) to another stationary point $a=q$ (corresponding to the static black hole state). Some constants $g^2\,,\tilde{\alpha}\,,q$ have been set to unity.}}
\label{fig2}\end{figure}

\subsection{Case 2: $k_0= 2\sqrt{\ft{2(n-1)}{n-2}}$}

In this case\footnote{When $k_0$ taking this value, our minimal theory (with $\mu=(n-2)/2$) can be conformally transformed into a non-minimal coupled model, which was first studied in \cite{Fan:2015tua}.}, $\gamma=-2$ and Eq.(\ref{aeq2}) can be integrated. The resulting first order equation is
\be \dot{a}+\tilde{\alpha}a^2\log{(\ft{a}{q})}=0 \,.\label{afirst}\ee
This equation can be further solved in terms of an exponential integral function,
\be \mathrm{Ei}\Big(\log{(\ft qa)}\Big)=-\tilde{\alpha}q u \,.\label{sol2}\ee

\begin{figure}[ht]
\begin{center}
\includegraphics[width=230pt]{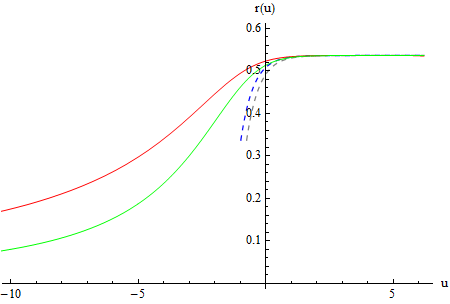}
\end{center}
\caption{{\it The evolution of local event horizon (red, blue) and apparent horizon (green, gray) for $n=4$ dimension. The dashed lines are plots for the Taylor series of $e^{-\tilde{\alpha} q u}$ (to second order). Some constants $g^2\,,\tilde{\alpha}\,,q$ have been set to unity.}}
\label{fig3}\end{figure}
We plot $a(u)$ as a function of the advanced time $u$ in Fig.\ref{fig2}. It is interesting to note that the system evolves from $u=-\infty$ to $u=+\infty$ and both points are stationary, namely $\dot{a}(\pm\infty)=0$. This is clear from the first order equation (\ref{afirst}). In fact, at $u=-\infty$ the space-time is pure AdS vacuum, which is linearly stable against perturbation\footnote{This is understood in the sense that the scalar field satisfies the BF bound. However,
for our dynamical solutions the perturbation of the scalar field is non-normalizable. In this case, the stability of the AdS vacua is not clear.}. Near the past infinity $u\rightarrow -\infty$, $a(u)$ behaves as $a\log{a}\sim -1/(\tilde{\alpha}u)$, indicating that the scalar just provides a weak source for the space-time evolution. However, after sufficiently long time, some non-linear effect becomes powerful to push the the space-time evolving into a stationary black hole state at the future infinity $u\rightarrow +\infty$.

We also find that as $u$ goes to $+\infty$, the system approaches the equilibrium at the rate of $e^{-u/u_0}$, where $u_0=1/(\tilde{\alpha}q)$ is the characteristic relaxation time.
The scalar function $a(u)$ and the Vaidya mass behave as
\be a(u)=q\Big(1-c_0e^{-\tilde{\alpha}q u}+\ft 32 c_0^2e^{-2\tilde{\alpha}q u}+\cdots \Big)\,,\quad M(u)=M_0\Big(1-\ft 12(n-1)^2c_0^2e^{-2\tilde{\alpha}q u}+\cdots  \Big)  \,,\ee
where $c_0$ is an integration constant which can be absorbed by constant shift of $u$. Formally, the local event horizon and apparent horizon also behave as Eq.(\ref{horizon}). In Fig.\ref{fig3}, we plot both of them as a function of the advanced time $u$ for $n=4$ dimension. It is clear that their evolution during the formation follows the basic features which were demonstrated in the first example.

Finally, we remark that this special case provides an analytical example how the linearly stable AdS vacua undergoes non-linear instability and spontaneously evolves into a stationary black hole state. We shall also point out that since there is energy input from the AdS infinity for our dynamical solutions, the non-linear instability discussed in this subsection is significantly different from the ordinary discussions of the non-linear instability of the AdS space-time\cite{Bizon:2011gg,Buchel:2012uh,Wu:2013qi,Buchel:2013uba,Bhattacharyya:2009uu}, in which no energy flux at infinity was considered.

\subsection{General case}

  Without loss of generality, we will set $k_0>0$ in this subsection. For generic $k_0$, the dynamical evolution equation Eq.(\ref{aeq2}) can also be integrated, giving rise to
 \be \dot{a}+\ft{\tilde{\alpha}}{\gamma+2}\,a^2\Big(1-\ft{q^{\gamma+2}}{a^{\gamma+2}}\Big)=0\,. \label{firstgene}\ee
 It follows that $\dot{a}=0$ for $a=q$ and $\dot{a}>0$, for $a<q$. This is valid for both $\gamma<-2$ and $\gamma>-2$ cases (note that the $\gamma=-2$ and $\gamma=0$ cases have been separately discussed in the above two subsections). Hence, $a=q$ is the stationary point corresponding to the final black holes state. For the $a=0$ point, the thing becomes more subtle. For lower lying $k_0$ satisfying $k_0<k_c$, where $k_c= 2\sqrt{\ft{2(n-3)}{n-2}}$ is a critical value, $\gamma$ becomes positive and $\dot a$ diverges in the $a\rightarrow 0$ limit. This implies that $a=0$ is not
 a stationary point and the space-time is unstable against perturbation. In fact, the global structure of the space-time around this point is similar to the $k_0=k_c$ case: the space-time becomes pure AdS except the $r=0\,,u=0$ point, which is an intrinsic singularity. On the contrary, for the higher value $k_0>k_c$, $\gamma$ becomes negative and $\dot a$ vanishes as $a$ approaches zero, indicating that $a=0$ is also a stationary point which corresponds to the AdS vacua state.
 \begin{figure}[ht]
\begin{center}
\includegraphics[width=210pt]{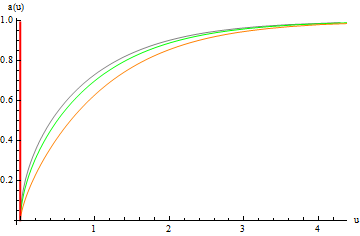}
\end{center}
\caption{{\it The plots of $a(u)$ for some lower lying $k_0$: $k_0=1/2$ (gray), $k_0=1$ (green), $k_0=3/2$ (orange). The vertical line denotes the singular limit of $\dot a$ as $a$ vanishes. Some constants $\tilde{\alpha}\,,q$ have been set to unity.}}
\label{fig4}\end{figure}

\begin{figure}[ht]
\begin{center}
\includegraphics[width=180pt]{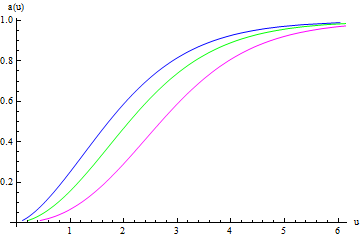}
\includegraphics[width=180pt]{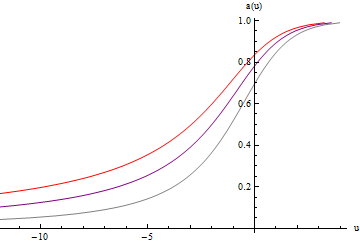}
\end{center}
\caption{{\it The plots of $a(u)$ for some higher value of $k_0$. In the left plot, $k_0=2.3$ (blue), $k_0=2.4$ (green), $k_0=2.5$ (orange). In the right plot, $k_0=3.5$ (gray), $k_0=4$ (purple), $k_0=4.5$ (red). Some constants $\tilde{\alpha}\,,q$ have been set to unity.}}
\label{fig5}\end{figure}

 The first order equation Eq.(\ref{firstgene}) can be analytically solved. We find
 \be \tilde{\alpha}q (u-u_i)=\fft{\gamma+2}{\gamma+1}\Big(\fft{a}{q}\Big)^{\gamma+1}
 {}_2F_1\Big(1\,,\ft{\gamma+1}{\gamma+2}\,,\ft{2\gamma+3}{\gamma+2}\,,\ft{a^{\gamma+2}}{q^{\gamma+2}}\Big)\,. \label{solgene}\ee
 Here $u_i$ is an integration constant and we set $u_i=0$ for $\gamma\geq -2$ and $u_i=i \pi$ for $\gamma<-2$ case\footnote{When $\gamma<-2$, the hypergeometric function becomes complex. Its imaginary part can be removed by setting $u_i=i\pi$ which guarantees the reality of the advanced time $u$ and $a$.}. In general, $a(u)$ turns out to be an implicit function of the advanced time $u$. In Fig.\ref{fig4}, we plot $a(u)$ as a function of the advanced time $u$ for some lower lying $k_0$ for the $n=4$ dimension. We see that their dynamical evolution follows the $k_0=k_c$ case except the singular limit of $\dot a$ at the $u=0$ point. In Fig.\ref{fig5}, We plot $a(u)$ for some higher value of $k_0$.
 We find that $\dot a$ vanishes in the $a\rightarrow 0$ limit for all the values $k_0>k_c$. In particular, for a range of parameter $-1<\gamma<0$, or equivalently $k_c<k_0<2\sqrt{2}$, the $a=0$ point was taken at some finite $u$ (which has been set to zero), instead of the past infinity. This can also be easily seen from the solutions Eq.(\ref{solgene}).

For all these solutions, the space-time exponentially approaches the equilibrium state at the future infinity. The relaxation time is universal, given by $u_0=1/(\tilde{\alpha}q)$. We find
\bea &&a(u)=q\Big(1-c_0e^{-\tilde{\alpha}q u}+\ft{1}{16}\big( (n-2)k_0^2-8(n-4)\big)c_0^2 e^{-2\tilde{\alpha}q u}+\cdots  \Big)\,,\nn\\
    &&M(u)=M_0\Big(1-\ft{1}{16}(n-2)(n-1)k_0^2c_0^2 e^{-2\tilde{\alpha}q u}+\cdots  \Big)\,. \eea
The dynamical evolution of both local event horizon and apparent horizon follows the basic properties which were shown in Fig.\ref{fig1} and Fig.\ref{fig3}.

\section{Conclusions}
In this paper, we study Einstein gravity minimally coupled to a scalar field with a given potential in general dimensions. We obtain large classes of static planar black holes with scalar hair which are asymptotic to AdS space-times. These hairy planar black holes contain only one free integration constant $q$ which is associated with the scalar field. Some physical quantities such as the mass, temperature and the entropy are all determined by functions of $q$. We discuss their global properties and derive their thermodynamic first laws.

For a particular case $\mu=(n-2)/2$, our minimal theory admits exact dynamic solutions in Eddington-Finkelstein like coordinates. These dynamical solutions describe the formation of black holes. They can be roughly classified into two different classes, depending on one of their parameters $k_0$ value. For the low lying $k_0$, satisfying $k_0\leq k_c$, where $k_c$ is some critical value, the dynamical evolution starts at $u=0$ at which the space-time becomes pure AdS but with a naked singularity at the origin. Hence, the space-time is unstable against perturbation and inevitably evolves into stationary black holes at the future infinity. For larger value $k_0>k_c$, the evolution starts from pure AdS vacua at $u=0$ or at the past infinity which is linearly stable. However, after sufficiently long time the space-time will also evolve into
stationary black hole states driven by some non-linear effects. Corresponding to these two classes solutions, we also present two exact examples.

In summary, our dynamic solutions provide more explicit and analytical examples describing the formation of planar black holes with scalar hair.

\section*{Acknowledgments}
The authors thank Prof. Hong Lu for valuable discussions. This work was in part supported by NSFC Grants No.~11275010, No.~11335012 and No.~11325522.


\begin{thebibliography}{99}

\bibitem{Balasubramanian:2010ce}
  V.~Balasubramanian, A. Bernamonti, J. de Boer, N. Copland, B. Craps, E. Keski-Vakkuri, B. M\"{u}ller, A. Sch\"{a}fer, M. Shigemori, and W. Staessens,
  {\it Thermalization of Strongly Coupled Field Theories,}
  Phys.\ Rev.\ Lett.\  {\bf 106}, 191601 (2011)
  [arXiv:1012.4753 [hep-th]].

\bibitem{Balasubramanian:2011ur}
 V.~Balasubramanian, A. Bernamonti, J. de Boer, N. Copland, B. Craps, E. Keski-Vakkuri, B. M\"{u}ller, A. Sch\"{a}fer, M. Shigemori, and W. Staessens,
  {\it Holographic Thermalization,}
  Phys.\ Rev.\ D {\bf 84}, 026010 (2011)
  [arXiv:1103.2683 [hep-th]].


\bibitem{Bizon:2011gg}
  P.~Bizon and A.~Rostworowski,
{\it On weakly turbulent instability of anti-de Sitter space,}
  Phys.\ Rev.\ Lett.\  {\bf 107}, 031102 (2011)
  [arXiv:1104.3702 [gr-qc]].

\bibitem{Buchel:2012uh}
  A.~Buchel, L.~Lehner and S.L.~Liebling,
{\it Scalar collapse in AdS,}
  Phys.\ Rev.\ D {\bf 86}, 123011 (2012)
  [arXiv:1210.0890 [gr-qc]].

\bibitem{Wu:2013qi}
  B.~Wu,
{\it On holographic thermalization and gravitational collapse of tachyonic scalar fields,}
  JHEP {\bf 1304}, 044 (2013)
  [arXiv:1301.3796 [hep-th]].

\bibitem{Buchel:2013uba}
  A.~Buchel, S.L.~Liebling and L.~Lehner,
{\it Boson stars in AdS spacetime,}
  Phys.\ Rev.\ D {\bf 87}, no. 12, 123006 (2013)
  [arXiv:1304.4166 [gr-qc]].


\bibitem{Bhattacharyya:2009uu}
  S.~Bhattacharyya and S.~Minwalla,
{\it Weak field black hole formation in asymptotically AdS spacetimes,}
  JHEP {\bf 0909}, 034 (2009)
  [arXiv:0904.0464 [hep-th]].


\bibitem{Zhang:2014sta}
  X.~Zhang and H.~L\"u,
{\it Exact black hole formation in asymptotically (A)dS and flat spacetimes,}
  Phys.\ Lett.\ B {\bf 736}, 455 (2014)
  [arXiv:1403.6874 [hep-th]].



\bibitem{Zhang:2014dfa}
  X.~Zhang and H.~L\"u,
{\it Critical behavior in a massless scalar field collapse with self-interaction potential,}
  Phys.\ Rev.\ D {\bf 91}, no. 4, 044046 (2015)
  [arXiv:1410.8337 [gr-qc]].


\bibitem{Lu:2014eta}
  H.~L\"u and X.~Zhang,
{\it Exact collapse solutions in $D = 4, \mathcal{N} = 4$ gauged supergravity and their generalizations,}
  JHEP {\bf 1407}, 099 (2014)
  [arXiv:1404.7603 [hep-th]].


\bibitem{Xu:2014xqa}
  W.~Xu,
{\it Exact black hole formation in three dimensions,}
  Phys.\ Lett.\ B {\bf 738}, 472 (2014)
  [arXiv:1409.3368 [hep-th]].


\bibitem{Ayon-Beato:2015ada}
  E.~Ay\'{o}n-Beato, M.~Hassa\"{\i}ne and J.~A.~M\'{e}ndez-Zavaleta,
  {\it (Super-)renormalizably dressed black holes,}
  Phys.\ Rev.\ D {\bf 92}, no. 2, 024048 (2015)
  [arXiv:1506.02277 [hep-th]].

\bibitem{Fan:2015tua}
  Z.~Y.~Fan and H.~Lu,
  {\it Static and Dynamic Hairy Planar Black Holes,}
  Phys.\ Rev.\ D {\bf 92}, no. 6, 064008 (2015)
  [arXiv:1505.03557 [hep-th]].

\bibitem{Henneaux:2002wm}
  M.~Henneaux, C.~Martinez, R.~Troncoso and J.~Zanelli,
  {\it Black holes and asymptotics of 2+1 gravity coupled to a scalar field,}
  Phys.\ Rev.\ D {\bf 65}, 104007 (2002)
  [hep-th/0201170].


\bibitem{Martinez:2004nb}
  C.~Martinez, R.~Troncoso and J.~Zanelli,
  {\it Exact black hole solution with a minimally coupled scalar field,}
  Phys.\ Rev.\ D {\bf 70}, 084035 (2004)
  [hep-th/0406111].


\bibitem{Anabalon:2012dw}
  A.~Anabalon,
{\it Exact hairy black holes,}
  Springer Proc.\ Phys.\  {\bf 157}, 3 (2014)
  [arXiv:1211.2765 [gr-qc]].

\bibitem{Gonzalez:2013aca}
  P.A.~Gonz\'alez, E.~Papantonopoulos, J.~Saavedra and Y.~V\'asquez,
{\it Four-dimensional asymptotically AdS black holes with scalar Hair,}
  JHEP {\bf 1312}, 021 (2013)
  [arXiv:1309.2161 [gr-qc]].

\bibitem{Anabalon:2013sra}
  A.~Anabalon and D.~Astefanesei,
 {\it On attractor mechanism of $AdS_{4}$ black holes,}
  Phys.\ Lett.\ B {\bf 727}, 568 (2013)
  [arXiv:1309.5863 [hep-th]].

\bibitem{Acena:2013jya}
  A.~Acena, A.~Anabalon, D.~Astefanesei and R.~Mann,
{\it Hairy planar black holes in higher dimensions,}
  JHEP {\bf 1401}, 153 (2014)
  [arXiv:1311.6065 [hep-th]].

\bibitem{Feng:2013tza}
  X.H.~Feng, H.~L\"u and Q.~Wen,
{\it Scalar hairy black holes in general dimensions,}
  Phys.\ Rev.\ D {\bf 89}, no. 4, 044014 (2014)
  [arXiv:1312.5374 [hep-th]].

\bibitem{Zloshchastiev:2004ny}
  K.G.~Zloshchastiev,
{\it On co-existence of black holes and scalar field,}
  Phys.\ Rev.\ Lett.\  {\bf 94}, 121101 (2005)
  [hep-th/0408163].


\bibitem{Fan:2015oca}
  Z.~Y.~Fan and H.~Lu,
  {\it Charged Black Holes with Scalar Hair,}
  JHEP {\bf 1509}, 060 (2015)
  [arXiv:1507.04369 [hep-th]].


\bibitem{Liu:2015tqa}
  H.~S.~Liu, H.~Lu and C.~N.~Pope,
  {\it Generalized Smarr formula and the viscosity bound for Einstein-Maxwell-dilaton black holes,}
  Phys.\ Rev.\ D {\bf 92}, no. 6, 064014 (2015)
  [arXiv:1507.02294 [hep-th]].

\bibitem{Feng:2015oea}
  X.~H.~Feng, H.~S.~Liu, H.~L¨¹ and C.~N.~Pope,
  {\it Black Hole Entropy and Viscosity Bound in Horndeski Gravity,}
  JHEP {\bf 1511}, 176 (2015)
  [arXiv:1509.07142 [hep-th]].

\bibitem{Hayward1}
  S.~A.~Hayward,
  {\it General laws of black hole dynamics,}
  Phys.\ Rev.\ D {\bf 49}, 6467 (1994).

\bibitem{Hayward2}
 S.~A.~Hayward,
  {\it Gravitational waves, black holes and cosmic strings in cylindrical symmetry,}
  Class.\ Quant.\ Grav.\  {\bf 17}, 1749 (2000).

\bibitem{Ida}
 D.~Ida,
  {\it No black hole theorem in three-dimensional gravity,}
  Phys.\ Rev.\ Lett.\  {\bf 85}, 3758 (2000).

\bibitem{Wang:2003bt}
  A.~Wang,
  {\it No-Go Theorem in Spacetimes with Two Commuting Spacelike Killing Vectors,}
  Gen.\ Rel.\ Grav.\  {\bf 37}, 1919 (2005)


\bibitem{Zhao:1992ad}
  Z.~Zhao and X.~X.~Dai,
  {\it A New method dealing with Hawking effects of evaporating black holes,}
  Mod.\ Phys.\ Lett.\ A {\bf 7}, 1771 (1992).


\bibitem{Zhao:1995jr}
  Z.~Zhao, J.~H.~Zhang and J.~Y.~Zhu,
  {\it Quantum thermal effect of arbitrarily accelerating Kinnersley black hole,}
  Int.\ J.\ Theor.\ Phys.\  {\bf 34}, 2039 (1995)
  [Int.\ J.\ Mod.\ Phys.\ A {\bf 20}, 1353 (2005)].


\end{thebibliography}
\end{document}